\documentstyle[12pt]{article}
\begin{document}
\baselineskip=24pt
\begin{center}
{\Large{\bf Evidence for Bound and Free Water Species
in the Hydration Shell of an Aqueous Micelle}}

\vspace{1cm}

{\large{\bf Sundaram Balasubramanian$^{a}$\footnote{email:bala@jncasr.ac.in}$^{*}$, Subrata Pal$^{b}$,
and Biman Bagchi$^{b}$\footnote{email:bbagchi@sscu.iisc.ernet.in}$^{*}$}}

\vspace{1cm}
$^{a}$ Chemistry and Physics of Materials Unit, \\
Jawaharlal Nehru Centre for Advanced Scientific Research,\\
Jakkur, Bangalore 560064, India.\\
\vspace{0.1cm}
$^{b}$Solid State and Structural Chemistry Unit,
Indian  Institute of Science,\\
Bangalore 560012, India.\\ 
\end{center}

\vspace{1cm}
            
\begin{center}
{\large {
\bf Abstract}}
\end{center}

{\bf Our atomistic molecular dynamics simulations reveal
the existence of bound and free water molecules in the 
hydration layer of an aqueous micelle. The bound water molecules can be either
singly or doubly hydrogen bonded to the polar head group on the surface of the micelle.
The ratio of bound to free water is found to be approximately equal to
9:1 at 300~K.}

\newpage
\section {Introduction}

 Water inevitably present at the surface of biological macromolecules and
 self-organized assemblies plays a critical role in the structure, stability
 and function of these systems~\cite{nandi00,biopapers}. However, the layer of water that surrounds
 these systems is rather thin, typically 1-3 layers thick. Thus, the study
 of hydration layer has turned out to be rather difficult. Dielectric
 relaxation studies measure the collective response of the whole system
 and, therefore, are not a sensitive probe of the dynamics of the hydration
 water. NMR techniques (NOE and NMRD) have the required spatial resolution
 but lack the dynamic resolution. Neutron scattering techniques are beginning to 
 be applied to study the dynamics of hydration layer in these systems~\cite{ruffle02}.
Solvation dynamics may be a reasonably good probe because it can have both the temporal
 and spatial resolution~\cite{sarkar96,lang99,riter98}, yet it provides only a collective 
response. In this circumstance, computer simulation can
 play a very important role in understanding the nature of the hydration water.
 
 As the hydration layer is spatially heterogeneous even on a molecular
 length scale and because the microscopic interactions are quite complex, a purely 
 analytical study of this system is 
 prohibitively difficult. In order to capture some of the basic physics
 of the dynamics in such systems, a model  in
 terms of a dynamic exchange between bound and free water 
molecules~\cite{nandi_bagchi_jpcA} has been proposed. The bound
 water molecules are those which are singly or doubly hydrogen bonded to
 the protein or to the surface of a self assembled aggregate while the free water molecules
are not. The microscopic dynamical event is an exchange between these two states of water 
within the hydration layer. The model predicts the emergence of a slow
decay when the binding energy is high. In this limit, the time constant of
the slow decay is just the inverse of the rate of bound to free 
transition\cite{nandi_bagchi_jpcA}.
Although this model has been
 semi-quantitatively successful in explaining wide range of behavior, the
 basic assumption of the existence of bound and free water molecules
 remained unsubstantiated.

  In addition to the above, bound and free water molecules give a convenient way to categorize the water molecules in the hydration layer. The bound molecules can be further sub-divided into two categories -- singly
  hydrogen bonded or doubly hydrogen bonded. We shall denote these two by IBW1
  and IBW2, respectively. Free water molecules are denoted by IFW (interfacial
  free water).

  Recently, we have presented several studies aimed at understanding
  various aspects of interfacial water~\cite{everything}. These studies were based
  on detailed atomistic molecular dynamics (MD) simulations of an anionic
  micelle, CsPFO (Cesium pentafluorooctanoate). These studies have 
  confirmed the existence of slow water and ion dynamics in the interfacial region. We have also studied the lifetime of hydrogen bonds that the water
molecules form with the micellar polar head groups (PHG) and found that
it becomes considerably longer than that between two water
molecules in the bulk.

   In the present work, we have extended our previous study to investigate
   in detail, the equilibrium structure of the interfacial water. 
   Our study has clearly revealed (we believe for the first time)
   that the water at the interface of CsPFO consists of 
   three different species -- IBW2, IBW1 and IFW, in the ratio, 1.1:8:0.9, i.e.,
   the bound to free water ratio is 9:1. This large ratio is a signature
   of the highly polar character of the anionic micelle.

\section{Results}

   As the details of the simulation have been discussed elsewhere~\cite{everything}, 
   we directly proceed to the discussion of the results. Figure 1
   illustrates the average geometries adopted by IBW1 and IBW2
   water species. The figure is not just an illustration -- we have
   given all the details about average bond lengths and the bond angles {\it calculated}
   from the simulation.  There are features which are absent in IBW1 but present in IBW2,
   such as a well-defined distance between the water and the second
   (non-bonded) oxygen of the polar head group. This feature at 4.7\AA~is prominent
   for IBW2 water molecules but absent for IBW1. An examination of the full distribution  
   of bond lengths and bond angles, and not just their averages, show that the 
   environment around bound water species is, in general, more well defined relative
   a water molecule in pure water.

    In figure 2 we present the monomer energy distribution, for all the
    three species, as also for water molecules in the bulk. It is seen that
    the interfacial water molecules have peaks at lower energies -- the
    doubly bonded species (IBW2) have the lowest potential energy. It is also worth
    noting that the bound water molecules have considerably lower energy
    values than the the free water molecules. It is this enhanced stability
    which makes the bound species identifiable, even when they are transient
    because of the dynamic exchange between the free and bound species. A large part
   of this stabilisation comes from the stronger hydrogen bonds that bound water 
    molecules form with the surfactant head groups.

     In figure 3 we provide a schematic of the free energies of the three
     species, calculated from their average concentrations. Despite the reduced monomer energy
     arising out of two water-headgroup hydrogen bonds,
     the IBW2 state is less stable than the IBW1 state due to entropic considerations (less 
      number of suitable configurations).
     The reversible reactions 
     between these states of water on such a surface should determine
     the dynamical response of interfacial water.

   Note that figure 3 describes the free energy and not the binding  energy of
the three species. The binding energy can be inferred from figure 2. The
total binding energy of the IBW2 species is indeed larger than IBW1.

\section{Conclusions} 
    In conclusion we note that the existence of identifiable bound and free 
water molecules on the surface can indeed help in developing a phenomenological
description of dynamics of water at complex interfaces.  The 9:1 ratio 
obtained (for IBW and IFW) is expected to be typical for ionic micelles. However,
this ratio is bound to decrease substantially for proteins due to the existence of hydrophobic
and less polar amino acid groups in its surface. The water on the protein/membrane
surface is expected to play a critical role in the molecular recognition of 
hydrophobic patches by incoming ligands or drug molecules.
Work in this direction is under progress.

 This work is supported by grants from the Department of Science and Technology and 
the Council of Scientific and Industrial Research, to both SB (JNC) and BB (IISC.).

\newpage
\begin{center}
FIGURE CAPTIONS
\end{center}

Fig. 1: Schematic description of the environment around bound interfacial waters, (a) IBW1, and 
  (b) IBW2. Numerical values of the geometrical parameters are average values obtained from the MD run. 
  Water molecules and surfactant headgroups are rigid entities in the interaction model.
  PHGO denotes the oxygen atom of the polar head group of the surfactant, and PHGC denotes
 the carbon atom in the head group. WO and WH denote the oxygen and hydrogen atoms of 
   the interfacial water, respectively. The broken lines between PHGO and WH denote the hydrogen
   bond.

Fig. 2: Distribution of monomer energies of interfacial water molecules (solid lines) compared to 
 that of  bulk water (dashed line). Solid lines from right to left represent the data for IFW, IBW1,
   and IBW2 species, respectively.

Fig. 3: Schematic description of the free energy (solid line) and internal energy (dashed line) 
     profiles of the interfacial water species. The species
    are in dynamical equilibrium with themselves and with water present in the bulk region of the
    micellar solution. The reaction coordinate is arbitrary and does not imply any distance. Barrier
     heights too are arbitrary.
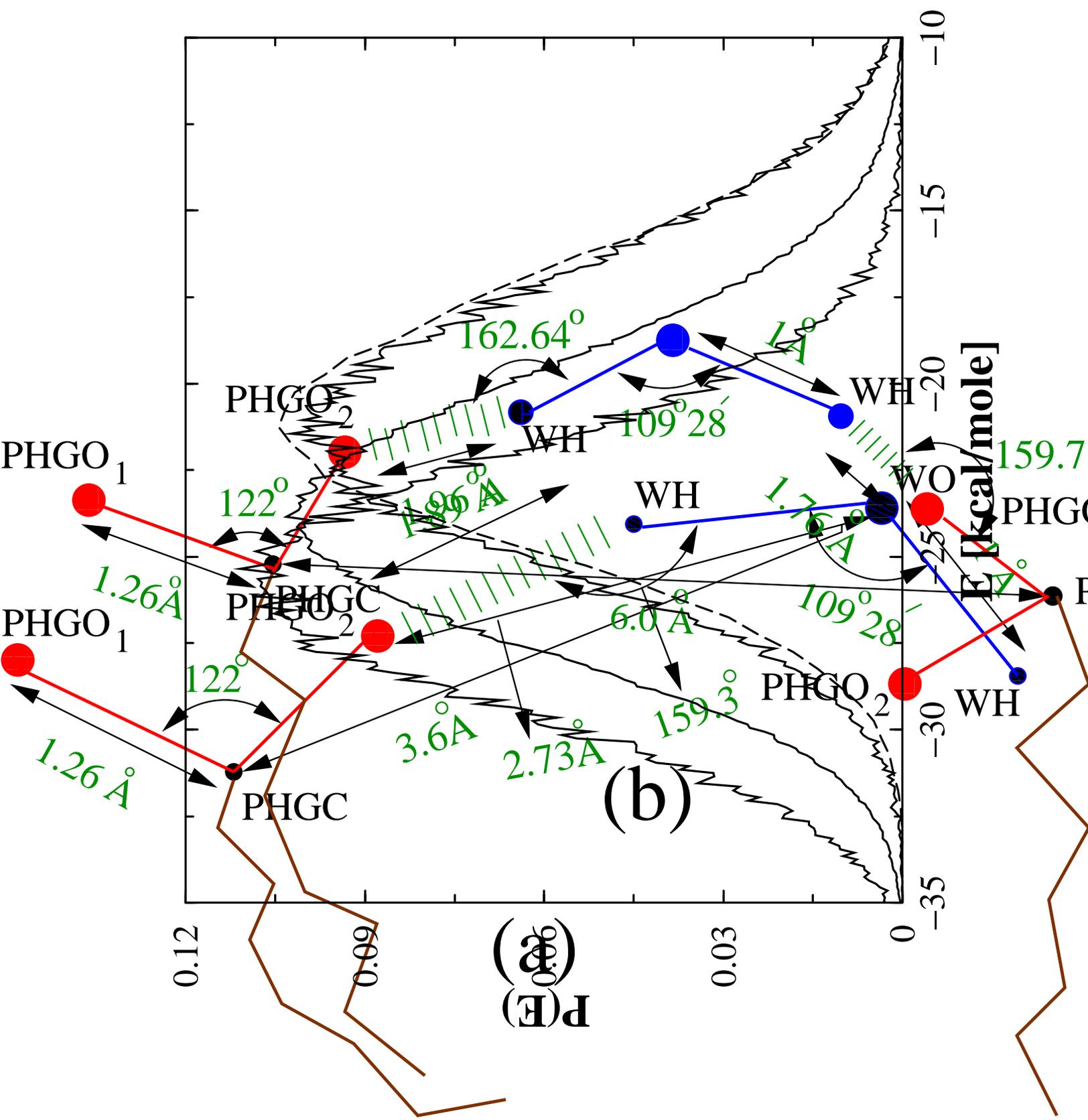
\end{document}